\documentclass{article}
\usepackage{graphicx}
\usepackage{newcent} 
\usepackage{fancyhdr}
\usepackage{hyperref} 
\usepackage{cite}
\usepackage{charter} 
\usepackage{multicol}
\usepackage{amsmath}

\usepackage[a4paper,top=3cm,bottom=3.3cm,left=3cm,right=3cm,footskip=1.5cm,truedimen,headheight=17pt]{geometry}
\hypersetup{colorlinks=true,citecolor=red,urlcolor=blue}
\begin{document}
\title{Efficient construction of many-body Fock states \\ having the lowest energies}%
\author{Andrzej Chrostowski and Tomasz Sowi\'nski \\[0.1cm]
           \small Institute of Physics of the Polish Academy of Sciences \\[-0.1cm]
           \small Al. Lotnik\'ow 32/46, 02-668 Warsaw, Poland}

\maketitle
\thispagestyle{fancy}

\begin{abstract}
To perform efficient many-body calculations in the framework of the exact diagonalization of the Hamiltonian one needs an appropriately tailored Fock basis built from the single-particle orbitals. The simplest way to compose the basis is to choose a finite set of single-particle wave functions and find all possible distributions of a given number of particles in these states. It is known, however, that this construction leads to very inaccurate results since it does not take into account different many-body states having the same energy on equal footing. Here we present a fast and surprisingly simple algorithm for generating the many-body Fock basis build from many-body Fock states having the lowest non-interacting energies. The algorithm is insensitive to details of the distribution of single-particle energies and it can be used for an arbitrary number of particles obeying bosonic or fermionic statistics. Moreover, it can be easily generalized to a larger number of components. Taking as a simple example the system of two ultra-cold bosons in an anharmonic trap, we show that exact calculations in the basis generated with the algorithm are substantially more accurate than calculations performed within the standard approach. \end{abstract}

\section{Introduction}
The exact diagonalization of the many-body Hamiltonian \cite{Lin,Lin2,Book2} is one of the simplest and straightforward methods of finding the ground state of the system of interacting quantum particles. It relies on a simple observation that having defined a finite set of ${\cal D}$ Fock states $|\mathtt{i}\rangle$ ($\mathtt{i}\in 1,\ldots,{\cal D}$) one can calculate all matrix elements of the Hamiltonian $\hat H$ of the system, $H_{ij}=\langle \mathtt{i}|\hat{H}|\mathtt{j}\rangle$, and numerically diagonalize corresponding matrix. As a result one obtains approximate decomposition of system's eigenstates into defined Fock basis $|\Psi_i\rangle= \sum_i \alpha_i |\mathtt{i}\rangle$ and their eigenenergies $E_i$. Obviously, the accuracy of this straightforward method strongly dependents not only on the number of Fock states used for calculations but also on a particular method the states are chosen from the infinite set of all possible Fock states. 

In typical physical scenario the total Hamiltonian of the many-body system can be divided to the sum of single-particle Hamiltonians describing dynamics of  particles confined in an external potential $V(\boldsymbol{r})$ and the remaining part describing mutual interactions between particles, {\it i.e.}, the many-body Hamiltonian of $N$ particles has a form
\begin{equation}
\hat{H} = \sum_{i=1}^N\left[-\frac{\hbar^2}{2m}\frac{\partial^2}{\partial\boldsymbol{r}_i^2}+V(\boldsymbol{r}_i)\right] + \hat{H}_{int}.
\end{equation}
Here we do not specify what is the form of the interaction Hamiltonian $\hat{H}_{int}$. In the most general case, it nontrivially depends on all particles' positions $\{\boldsymbol{r}_1,\ldots,\boldsymbol{r}_N\}$ and their derivatives. In the following, we assume that the single-particle part of the Hamiltonian is already diagonalized and all its eigenstates $\varphi_k(\boldsymbol{r})$ and their eigenenergies $\varepsilon_k$ (sorted along ascending order) are known. Having these states, it is very convenient to consider many-body Hilbert space of $N$ particles as spanned by Fock states build from these single-particle orbitals. It means that any Fock state can be written formally in the second quantization formalism as a sequence of numbers of particles occupying individual orbitals
\begin{equation} \label{2ndQuant}
|\mathtt{i}\rangle = |n_0\,n_1 \ldots\rangle,
\end{equation}
with $\sum_k n_k = N$. Of course, in the case of fermionic particles, an additional constraint has to be imposed assuring that for any $k$ occupation $n_k\in\{0,1\}$. In this way, quantum indistinguishability and statistics are taken into account. The non-interacting energy of this state is $E_{\mathtt{i}}=\sum_k n_k \varepsilon_k$. This notation immediately suggests one of the simplest methods of limiting the size of the Fock space and choosing only ${\cal D}$ of them as required by any numerical method based on diagonalization \cite{ED}. Namely, one limits the number of single-particle orbitals to some chosen cut-off ${\cal C}$, $k\in \{0,\ldots,{\cal C}\}$, and use them to build all possible Fock states for further calculations. In this approach, to increase the numerical accuracy of the many-body diagonalization, one slightly modifies a shape of single-particle orbitals \cite{Przemek} or increases the cut-off ${\cal C}$. Unfortunately, along with increasing cut-off ${\cal C}$ the number of Fock states grows exponentially (${\cal D}_\mathtt{C} = ({\cal C}+N)!/({\cal C}!\,N!)$ and ${\cal D}_\mathtt{C} = ({\cal C}+1)!/[({\cal C}-N+1)!\,N!]$ for bosons and fermions, respectively). In consequence, the method is naturally limited by numerical resources (mainly by available memory and computational time) and it is commonly viewed as very demanding from a computational perspective.

To limit exponential growth of the considered Hilbert space one can change the way the Fock basis is constructed. As discussed in the literature (see for example \cite{Haugset,Deuretz,Poland}), generating the Fock basis directly from the limited number of single-particle states is not the most efficient way of obtaining well-converged results, since then an energetic hierarchy in the Fock basis is completely neglected. Namely, this approach takes into account Fock states with relatively large energies fitting to small cut-offs $\cal C$ and neglects at the same time other Fock states with substantially lower energies but having larger single-particle excitations. As argued in \cite{Poland}, to increase the accuracy of the exact diagonalization without extending numerical efforts, one should select Fock states with the lowest non-interacting energy $E_{\mathtt{i}}=\sum_k n_k\varepsilon_k$ rather than states with the lowest excitations $k\leq {\cal C}$.

In general, selecting the Fock states $\{|\mathtt{i}\rangle\}$ having the lowest non-interacting energy is not a trivial task. Only in the case of equally distributed single-particle energies $\varepsilon_k-\varepsilon_{k-1}=\Delta$ (like in the one-dimensional parabolic confinement) it can be done quite easily since then the non-interacting energies can be represented uniquely by integer numbers. Consequently, the problem can be reduced to the task of finding different partitions of an integer (representing the non-interacting energy) into a fixed number of parts (representing individual particles) \cite{Book}. A detailed explanation is presented for example in \cite{Poland}. Unfortunately, a simple generalization of this approach to cases when single-particle energies are not equally distant does not exist. Therefore, one needs to apply a direct selection of states with the lowest non-interacting energy from a basis of states having the lowest excitations (determined by a single-particle cut-off ${\cal C}$). This, however, is very inefficient and consuming huge amounts of computational facilities (memory and computation time). Consequently, its usefulness is strongly limited. 

In this work, we present a simple numerical algorithm of generating the cropped Fock basis for $N$ particles (bosons as well as fermions) in the general case of any complete set of single-particle orbitals $\varphi_k(\boldsymbol{r})$ having ascending single-particle energies $\varepsilon_0\leq\varepsilon_1\leq\ldots$. As the input, the algorithm requires only one parameter, namely the maximal energy $\boldsymbol{E}$ of states in generated Fock basis. As a result one obtains a complete set of all ${\cal D}_\mathtt{E}$ Fock states having energy no larger than declared maximal energy $\boldsymbol{E}$. Importantly, an order of states in the generated set is well-known since automatically they are sorted lexicographically. Therefore, one can easily find not only $\mathtt{i}$-th Fock state but also perform reversed identification on demand and find the index $\mathtt{i}$ of the Fock state represented by its occupations. This property of the set is crucial for quick generation of a matrix representation of a many-body Hamiltonian matrix. A fundamental advantage of the algorithm presented is that Fock states are generated sequentially one-by-one in ${\cal O}({\cal D}_\mathtt{E})$ time, {\it i.e.}, an improvement with respect to the case of a direct selection method is tremendous.   
\section{The algorithm}
\begin{figure}[t!] 
\centering
\includegraphics[width=0.7\linewidth]{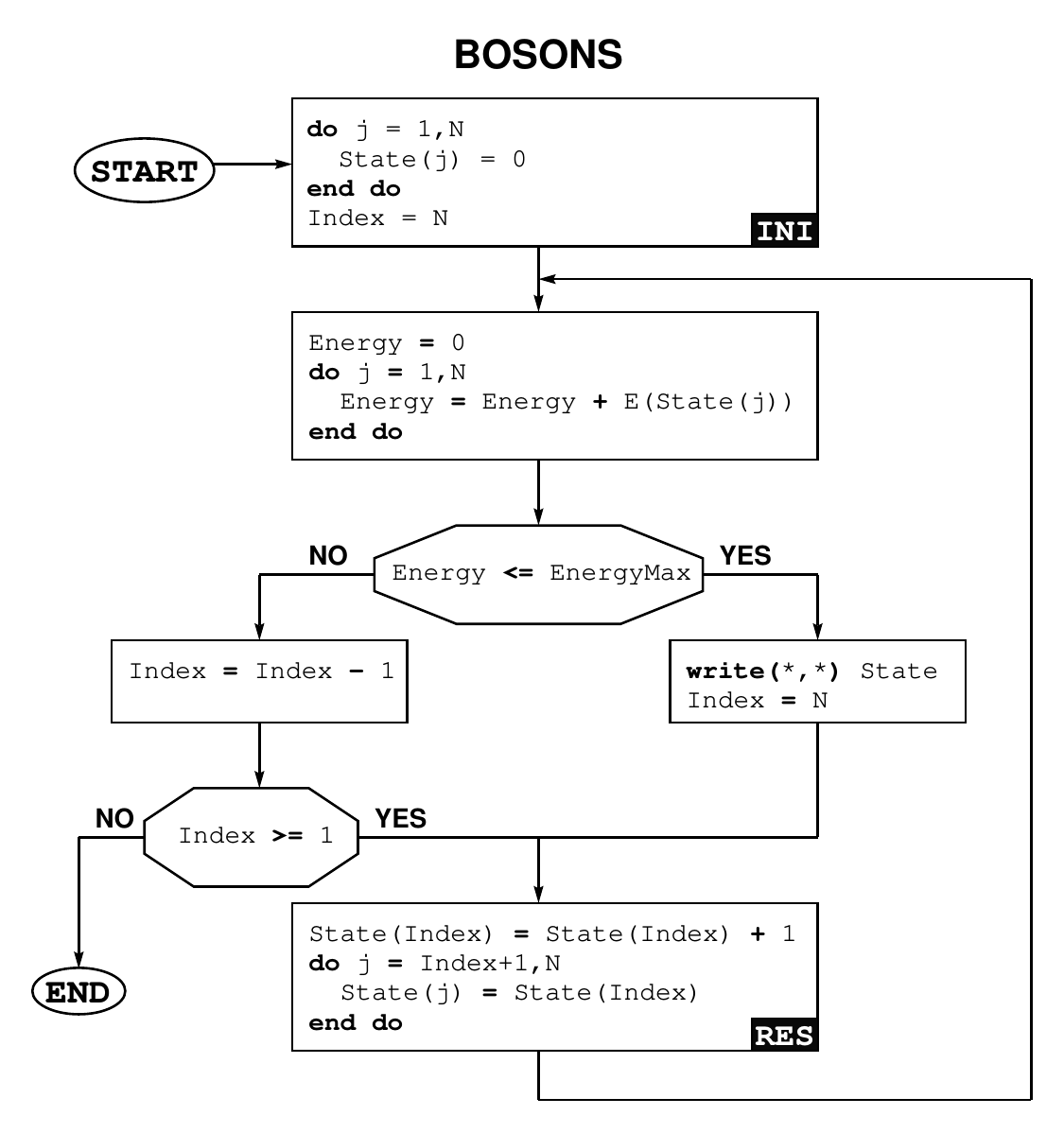}
\caption{Schematic flowchart of the algorithm generating a sequence of Fock states with the lowest non-interacting energy of $N$ bosons from a given set of single-particle orbitals having ascending energies $\varepsilon_k$ stored in variables \texttt{E(k)}. The maximal energy $\boldsymbol{E}$ is stored in \texttt{EnergyMax}. Initial and Reset boxes (\textbf{INI} and \textbf{RES}, respectively) are marked for convenient comparison with the corresponding algorithm for fermionic particles. \label{FigBosons}}
\end{figure}

Counterintuitively, the best way to present the algorithm for generating the cropped Fock basis of states with the lowest non-interacting energy is to represent Fock states in the first quantization formalism, {\it i.e.}, instead of the number of particles occupying single-particle orbitals one needs to remember single-particle states occupied by individual particles. It means that the Fock state $|\mathtt{i}\rangle$ is represented by a set of $N$ numbers $(i_1,\ldots,i_N)$ rather than a vector \eqref{2ndQuant} encoding occupations. The relation between the two is straightforward. As an example, the non-interacting ground-state of $N=4$ bosons in both notations reads $|\mathtt{1}\rangle=|4\,0\,\ldots\rangle=(0,0,0,0)$ while in the case of $N=4$ fermions it has a form $|\mathtt{1}\rangle=|1\,1\,1\,1\,0\ldots\rangle=(0,1,2,3)$. Since particles are indistinguishable, it is understood that the states encoded in the first quantization are appropriately 
(anti-)symmetrized and only for simplicity they are represented by sets with ascending numbers, $i_1\leq i_2\leq\ldots\leq i_N$.

\begin{table}[t!]
\centering
\begin{multicols}{1}
{\bf (a)} Harmonic oscillator ($\varepsilon_i = 0.5+i$)  \hfill \mbox{ }\\[0.1cm]
\begin{tabular}{|ccc|}
\hline
{\bf Index} & {\bf Occupations} & {\bf Energy} \\
\hline\hline 
$|\mathtt{ 1}\rangle$ & $|\mathtt{ 4 0 0 0 0 }\rangle$ & 2.000000 \\
$|\mathtt{ 2}\rangle$ & $|\mathtt{ 3 1 0 0 0 }\rangle$ & 3.000000 \\
$|\mathtt{ 3}\rangle$ & $|\mathtt{ 3 0 1 0 0 }\rangle$ & 4.000000 \\
$|\mathtt{ 4}\rangle$ & $|\mathtt{ 3 0 0 1 0 }\rangle$ & 5.000000 \\
$|\mathtt{ 5}\rangle$ & $|\mathtt{ 3 0 0 0 1 }\rangle$ & 6.000000 \\
$|\mathtt{ 6}\rangle$ & $|\mathtt{ 2 2 0 0 0 }\rangle$ & 4.000000 \\
$|\mathtt{ 7}\rangle$ & $|\mathtt{ 2 1 1 0 0 }\rangle$ & 5.000000 \\
$|\mathtt{ 8}\rangle$ & $|\mathtt{ 2 1 0 1 0 }\rangle$ & 6.000000 \\
$|\mathtt{ 9}\rangle$ & $|\mathtt{ 2 0 2 0 0 }\rangle$ & 6.000000 \\
$|\mathtt{10}\rangle$ & $|\mathtt{ 1 3 0 0 0 }\rangle$ & 5.000000 \\
$|\mathtt{11}\rangle$ & $|\mathtt{ 1 2 1 0 0 }\rangle$ & 6.000000 \\
$|\mathtt{12}\rangle$ & $|\mathtt{ 0 4 0 0 0 }\rangle$ & 6.000000 \\
\hline
\end{tabular} \\[0.2cm]
\vfill
\columnbreak
{\bf (b)} Anharmonic well defined by \eqref{anharm} \hfill \mbox{ }\\[0.1cm]
\begin{tabular}{|ccc|}
\hline
{\bf Index} & {\bf Occupations} & {\bf Energy} \\
\hline\hline 
$|\mathtt{ 1}\rangle$ & $|\mathtt{ 4 0 0 0 0 0 0 0 }\rangle$ & 2.000000 \\
$|\mathtt{ 2}\rangle$ & $|\mathtt{ 3 1 0 0 0 0 0 0 }\rangle$ & 3.000000 \\
$|\mathtt{ 3}\rangle$ & $|\mathtt{ 3 0 1 0 0 0 0 0 }\rangle$ & 3.800000 \\
$|\mathtt{ 4}\rangle$ & $|\mathtt{ 3 0 0 1 0 0 0 0 }\rangle$ & 4.440000 \\
$|\mathtt{ 5}\rangle$ & $|\mathtt{ 3 0 0 0 1 0 0 0 }\rangle$ & 4.952000 \\
$|\mathtt{ 6}\rangle$ & $|\mathtt{ 3 0 0 0 0 1 0 0 }\rangle$ & 5.361600 \\
$|\mathtt{ 7}\rangle$ & $|\mathtt{ 3 0 0 0 0 0 1 0 }\rangle$ & 5.689280 \\
$|\mathtt{ 8}\rangle$ & $|\mathtt{ 3 0 0 0 0 0 0 1 }\rangle$ & 5.951424 \\
$|\mathtt{ 9}\rangle$ & $|\mathtt{ 2 2 0 0 0 0 0 0 }\rangle$ & 4.000000 \\
$|\mathtt{10}\rangle$ & $|\mathtt{ 2 1 1 0 0 0 0 0 }\rangle$ & 4.800000 \\
$|\mathtt{11}\rangle$ & $|\mathtt{ 2 1 0 1 0 0 0 0 }\rangle$ & 5.440000 \\
$|\mathtt{12}\rangle$ & $|\mathtt{ 2 1 0 0 1 0 0 0 }\rangle$ & 5.952000 \\
$|\mathtt{13}\rangle$ & $|\mathtt{ 2 0 2 0 0 0 0 0 }\rangle$ & 5.600000 \\
$|\mathtt{14}\rangle$ & $|\mathtt{ 1 3 0 0 0 0 0 0 }\rangle$ & 5.000000 \\
$|\mathtt{15}\rangle$ & $|\mathtt{ 1 2 1 0 0 0 0 0 }\rangle$ & 5.800000 \\
$|\mathtt{16}\rangle$ & $|\mathtt{ 0 4 0 0 0 0 0 0 }\rangle$ & 6.000000 \\
\hline
\end{tabular}
\end{multicols}
\caption{Fock basis build form states with non-interacting energies not larger than $\boldsymbol E=6$ for the system of $N=4$ bosons obtained with the algorithm presented in Fig.~\ref{FigBosons} for two different single-particle energy spectra \eqref{HamDef} and \eqref{anharm}. In both cases obtained states are generated in a lexicographic order. Note that in both cases some of states have highly-excited particles while their non-interacting energy is relatively low.   See the main text for a discussion. \label{Table1}}
\end{table}

The idea of the algorithm is very similar for both quantum statistics. Therefore, let us first perform a short presentation for the case of $N$ bosons. Schematic flowchart of the algorithm (in the Fortran-like code) is presented in Fig.~\ref{FigBosons}. The input requirements are: number of particles $N$, the single-particle energies $\varepsilon_k$ sorted along ascending order (stored in \texttt{E(k)}), and the maximal energy $\boldsymbol{E}$ (stored in  \texttt{EnergyMax}). As an output one gets the series of Fock states $\{|\mathtt{i}\rangle\}$ represented by a set of numbers $(i_1,\ldots,i_N)$ (stored temporarily in \texttt{State(j)}) which are generated in lexicographical order. Example outputs of the algorithm for two different situations are presented in Table~\ref{Table1}. Table \ref{Table1}a presents resulting Fock basis for the one-dimensional harmonic confinement having equally distant single-particle energy levels
\begin{subequations} \label{HamDef}
\begin{align}
\varepsilon_0 &= 0.5, \\
\varepsilon_k &=\varepsilon_{k-1}+1, \,\,\mathrm{for}\,\,k>0
\end{align}
\end{subequations}
with the maximal energy $\boldsymbol{E}=6$. As it is seen, only twelve Fock states have this property (${\cal D}_\mathtt{E}=12$). At this point, one should note a huge reduction of the considered Hilbert space when compared to the standard approach of single-particle cut-off. It is clearly seen that in the case of the former method, to capture all states with non-interacting energy no larger than $\boldsymbol{E}=6$, one needs to use cut-off ${\cal C}=4$ (state $|\mathtt{5}\rangle$ has one particle excited to the single-particle state with energy $\varepsilon_4=4.5$). It means that the cropped Hilbert space generated with the standard cut-off approach would have a much larger dimension of ${\cal D}_\mathtt{C}=70$.

To show that the algorithm works also for other confinements in Table \ref{Table1}b we present the output for the same maximal energy $\boldsymbol{E}=6$, but for anharmonic confinement having decreasing energy distances between single-particle orbitals. As an example we set
\begin{subequations} \label{anharm}
\begin{align}
\varepsilon_0 &= 0.5, \\
\varepsilon_k &=\varepsilon_{k-1}+(4/5)^{k-1}, \,\,\mathrm{for}\,\,k>0.
\end{align}
\end{subequations}
It is clearly seen that in this case, the gain from the algorithm is even larger. There are only sixteen Fock states (${\cal D}_\mathtt{E}=16$) having appropriate energy, while some of them have highly excited particles. For example, the state $|\mathtt{8}\rangle$ has one particle with energy $\varepsilon_7=92\,991/15\,625$. It means that in the standard cut-off method one needs to introduce ${\cal C}=7$ to capture all the states and the resulting size of the Hilbert space would be considerably larger ${\cal D}_\mathtt{C}=1\,980$.

\begin{figure}[t!]
\centering
\includegraphics[width=0.7\linewidth]{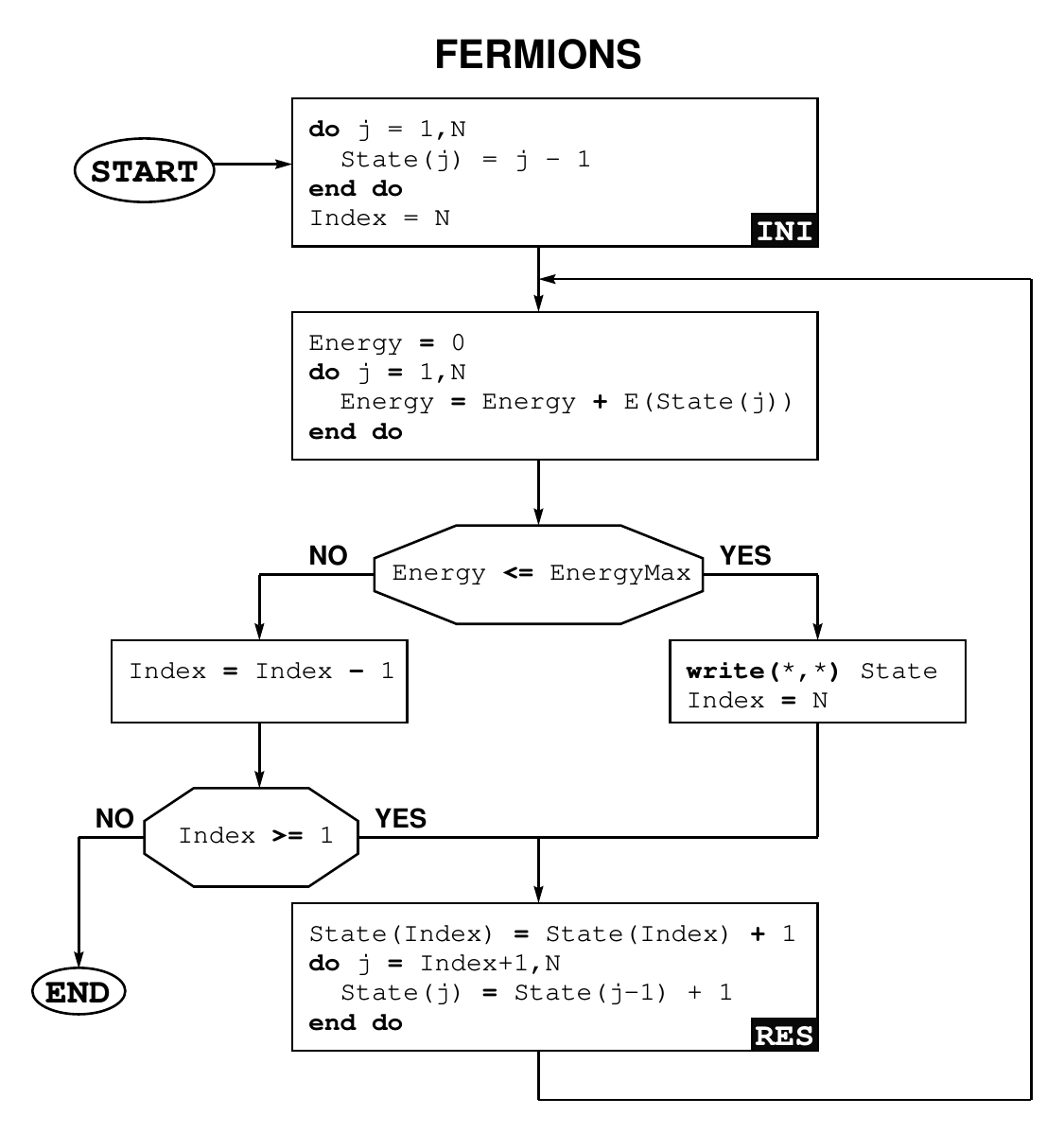}
\caption{Adaptation of the algorithm presented in Fig.~\ref{FigBosons} to the case of fermionic systems. To fulfill fermionic statistics forbidding any two particles to occupy the same single-particle state the Initial Block and the Reset Block need to be modified (\textbf{INI} and \textbf{RES}, respectively). Compare with corresponding blocks in Fig.~\ref{FigBosons}. \label{FigFermions}}
\end{figure}
In the case of fermionic statistics, any two particles cannot occupy the same single-particle orbital. Therefore, the algorithm needs to be modified to take into account this constraint. Fortunately, the changes are not substantial and they concern only the Initial Block and the Reset Block (marked in Fig.\ref{FigBosons} as \textbf{INI} and \textbf{RES}, respectively). In the Initial Block, the change is trivial and it is reduced to a simple change of the non-interacting ground state. In the Reset Block, particles having indexes larger than currently stored index should have consecutive occupations instead of occupations equal to the occupation of the indexed particle. For completeness, in Fig.~\ref{FigFermions} we present a schematic flowchart of the algorithm with these two changes directly incorporated. 

Finally, let us emphasize that the algorithm presented can be very easily generalized to cases of multicomponent mixtures of particles with different statistics. One may think that the simplest way of such generalization is to generate Fock basis for each component separately and then build a multicomponent basis by a simple tensor composition. However, this approach is not correct since the resulting Fock basis does not contain states with the lowest non-interacting energy. It is especially crucial when the numbers of particles in different components are different or when different components have different single-particle energy spectra.  To overcome this difficulty one should apply above algorithms to each component recursively taking care on the total non-interacting energy of generated Fock state, {\it i.e.}, when the state of a given component is generated (instruction \mbox{\tt write(*,*) State} in flowcharts) next algorithm for the following component should be executed. To assure that the non-interaction energy of the complete Fock state is bounded, the {\tt EnergyMax} variable for a given component should be set as the maximal non-interacting energy $\boldsymbol{E}$ lowered by a sum of current energies of components generated already. The complete Fock state is accepted as a member of the Fock basis in the last component output.

\section{Physical example}
\begin{figure}[t!]
\centering
\includegraphics[width=0.8\linewidth]{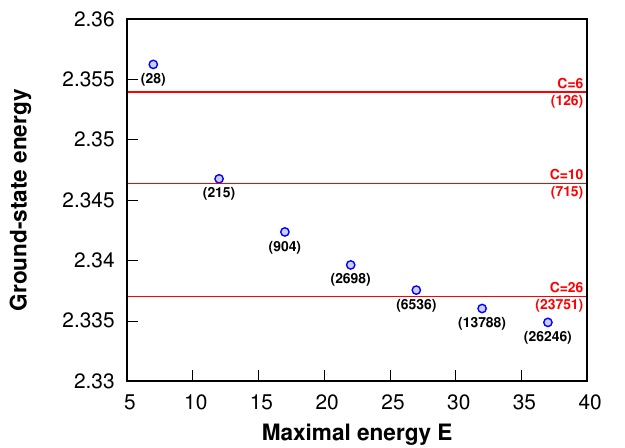}
\caption{The ground-state energy obtained for the system of $N=2$ bosons for different numerical schemes of generating the Fock basis. Blue points correspond to the case of exact diagonalization in the basis of Fock states with the lowest non-interacting energy, while red lines represent results when the standard cut-off method of the single-particle basis is used. The ground-state energy decreases along with the size of the Fock basis ${\cal D}$ (numbers in parenthesizes) determined by the maximal energy $\boldsymbol{E}$ and the cut-off ${\cal C}$, respectively. Note improved convergence for the same sizes of the Fock space when the Fock states with the lowest non-interacting energies are used for calculations.  \label{Fig3}}
\end{figure}

To show that the presented method of building the many-body Fock basis substantially increases  efficiency of the exact diagonalization method let us consider the simplest model of $N=2$ ultra-cold bosons confined in a one-dimensional anharmonic trap and interacting via contact forces. In the dimensionless units the Hamiltonian of the system reads
\begin{equation} \label{HamEx}
\hat{H} = \sum_{i=1}^N\left[-\frac{1}{2}\frac{\partial^2}{\partial{x}_i^2}+\frac{1}{2}|x_i|^{2-\alpha}\right] + g\sum_{i>j}\delta(x_i-x_j),
\end{equation}
where $\alpha$ controls an anharmonicity of the trap (for $\alpha=0$ the standard harmonic confinement is restored) and $g$ is the interaction strength between particles. In the following example we set $\alpha=0.5$ and $g=0.1$.

In Fig.~\ref{Fig3} we present the ground-state energy of the Hamiltonian \eqref{HamEx} obtained with two complementary methods. First, with blue points, we display the ground-state energy obtained when the Fock basis is generated with the presented algorithm. Different points correspond to different maximal energies $\boldsymbol{E}$ and they are labeled by corresponding sizes of obtained Fock basis ${\cal D}_\mathtt{E}$. For comparison, using red horizontal lines, we present the ground-state energies obtained with the standard method of truncating the single-particle basis. Different lines correspond to different cut-offs ${\cal C}$ and they are additionally labeled by corresponding sizes of the cropped many-body Hilbert spaces ${\cal D}_\mathtt{C}$. It is clearly seen that along with increasing sizes of the Fock basis the ground-state energies decreases (ground-state energies obtained with any approximate method are always bounded from below by the exact ground-state energy of the system). Note that, the standard cut-off method is very inefficient when it is compared to diagonalization in the basis of energetically the lowest Fock states. For a given $\cal C$ the same ground-state energy may be obtained for much smaller Hilbert spaces provided that one selects Fock states systematically according to the non-interacting energy. For example, the ground-state energy obtained for ${\cal C}=26$ corresponding to the size of the Hamiltonian matrix ${\cal D}_\mathtt{C}=23\,751$ is almost the same as the energy obtained for almost four times smaller Hamiltonian matrix (${\cal D}_\mathtt{E}=6\,536$) calculated in the basis of Fock states with non-interacting energy not larger than $\boldsymbol{E}=26$.

\section{Conclusions}
Accuracy of the exact diagonalization of many-body Hamiltonians crucially depends on an appropriate selection of the Fock states. Typically, they are chosen from the set of eigenstates of the non-interacting system. Then, the efficiency of the method can be significantly improved if, instead of states having particles with the lowest excitations, the states with the lowest non-interacting energies are chosen. In this work, we present a very efficient and surprisingly simple algorithm generating all Fock states with the lowest non-interacting energy from a given set of single-particle orbitals. In this way, we give the base for further efficient calculations of a variety of many-body scenarios. In principle, the algorithm can be adapted to any many-body system for which a non-interacting Hamiltonian can be written as a sum of independent single-particle Hamiltonians. Therefore, it can be extremely useful in many different physical scenarios. 
\vspace{6pt} 

\section*{Acknowledgments}
The authors would like to thank Przemys{\l}aw Ko\'scik for constructive comments. This work was supported by the (Polish) National Science Center through Grant No. 2016/22/E/ST2/00555.

\end{document}